\documentclass[superscriptaddress, preprint]{revtex4-1}

\usepackage{graphicx}
\usepackage{grffile}
\usepackage{MnSymbol}

\usepackage[utf8]{inputenc}
\usepackage{amsmath} 
\usepackage{braket}
\usepackage{float}
\usepackage[english]{babel}
\usepackage{tikz}
\usetikzlibrary {positioning}
\usetikzlibrary {fit}
\usetikzlibrary{calc}
\usetikzlibrary{decorations.pathreplacing}
\usetikzlibrary{shapes.multipart}
\usetikzlibrary{patterns}
\usepackage{marginnote}
\usepackage{comment}

\usepackage{url}

\begin{document}
\title{Deep neural networks for high harmonic spectroscopy in solids}

%\date{\today}

\begin{comment}
\author{Nikolai D Klimkin,\authormark{1, 2, *} \'Alvaro Jim\'enez-Gal\'an,\authormark{3} Rui E. F. Silva,\authormark{4} and Misha Ivanov\authormark{3, 5, 6}}

\address{\authormark{1}Russian Quantum Center, Bolshoy Bulvar 30, bld. 1, Moscow, Russia 121205\\
\authormark{2}Faculty of Physics, Lomonosov Moscow State University, Leninskie Gory 1-2, Moscow 119991
\authormark{3}Max-Born-Institute, Max-Born Stra{\ss}e 2A, D-12489 Berlin, Germany\\
\authormark{4}Dpto. F\'isica Te\'orica de la Materia Condensada, Universidad Aut\'onoma de Madrid, Madrid, Spain\\
\authormark{5}Department of Physics, Humboldt University, Newtonstra{\ss}e 15, D-12489 Berlin, Germany\\
\authormark{6}Blackett Laboratory, Imperial College London, South Kensington Campus, SW7 2AZ London, United Kingdom}

\email{\authormark{*}nd.klimkin@physics.msu.ru}
\end{comment}

\author{Nikolai D. Klimkin}
\email{nd.klimkin@physics.msu.ru}
\affiliation{Russian Quantum Center, Bolshoy Bulvar 30, bld. 1, Moscow, Russia 121205}
\affiliation{Faculty of Physics, Lomonosov Moscow State University, Leninskie Gory 1-2, Moscow 119991}
\author{\'Alvaro Jim\'enez-Gal\'an}
\affiliation{Max-Born-Institute, Max-Born Stra{\ss}e 2A, D-12489 Berlin, Germany}
\author{Rui E. F. Silva}
\affiliation{Dpto. F\'isica Te\'orica de la Materia Condensada, Universidad Aut\'onoma de Madrid, Madrid, Spain}
\author{Misha Ivanov}
\affiliation{Max-Born-Institute, Max-Born Stra{\ss}e 2A, D-12489 Berlin, Germany}
\affiliation{Department of Physics, Humboldt University, Newtonstra{\ss}e 15, D-12489 Berlin, Germany}
\affiliation{Blackett Laboratory, Imperial College London, South Kensington Campus, SW7 2AZ London, United Kingdom}

\begin{comment}
\footnotetext[1]
\footnotetext[2]
\footnotetext[3]
\footnotetext[4]{Dpto. F\'isica Te\'orica de la Materia Condensada, Universidad Aut\'onoma de Madrid, Madrid, Spain}
\footnotetext[5]{Department of Physics, Humboldt University, Newtonstra{\ss}e 15, D-12489 Berlin, Germany}
\footnotetext[6]{Blackett Laboratory, Imperial College London, South Kensington Campus, SW7 2AZ London, United Kingdom}
\end{comment}

\date{\today}

\begin{abstract}
%\begin{center}
Neural networks are a prominent tool for identifying and modeling complex patterns, which are otherwise hard to detect and analyze. While machine learning and neural networks have 
been finding applications across many areas of science and technology, their use in decoding ultrafast dynamics of quantum systems driven by strong laser fields 
has been limited so far. Here we use deep neural networks to analyze simulated noisy spectra of highly nonlinear optical response of a 2-dimensional gapped graphene crystal to intense few-cycle laser pulses. We show that a computationally simple 1-dimensional system provides a useful "nursery school" for our neural network, allowing it to be easily retrained to treat more complex systems, recovering the band structure and spectral phases of the incident few-cycle pulse with high accuracy, in spite of significant amplitude noise and phase jitter. Our results both offer a new tool for attosecond spectroscopy of quantum dynamics in solids and also open a route to developing all-solid-state devices for complete characterization of few-cycle pulses, including their nonlinear chirp and the carrier envelope phase. 
%\end{center}
\end{abstract}

\maketitle

Electrons provide the fundamental first step in response of matter to light. 
The feasibility of shaping light pulses at the scale of individual oscillations, from mid-IR to UV~\cite{Wirth195, Hassan2016}, offers rich opportunities for controlling electronic response to light on sub-cycle timescale (e.g.~\cite{schultze2013controlling, schiffrin2013optical, kelardeh2016attosecond, kelardeh2015graphene, motlagh2019topological, garg2016multi, lakhotia2020laser, vampa2015linking, reimann2018subcycle, langer2018lightwave, jimnezgaln2019lightwave}), leading to a variety of fascinating phenomena such as optically induced anomalous Hall effect~\cite{motlagh2020anomalous, sato2019light, mciver2020light}, topological phase transitions with polarization-tailored light~\cite{jimnezgaln2019lightwave}, or the topological resonance~\cite{motlagh2019topological}.
%On this time-scale, strong light fields exert forces comparable to internal forces acting on electrons in matter, lock the electronic response to light oscillations, and support electronic coherence. 
Over multiple laser cycles, control of electron dynamics with light also enables the so-called Floquet engineering -- the tailored modification of the cycle-average properties 
of a light-dressed system, 
see e.g.~\cite{oka2019floquet} for a recent review.

In this context, starting with the pioneering work~\cite{ghimire2011observation}, high harmonic spectroscopy  
has developed into a powerful tool for exploring laser-driven 
electron dynamics in solids, see e.g. recent reviews ~\cite{vampa2017merge, kruchinin2018colloquium, ghimire2019high}.
Examples include identification of the common physical mechanisms underlying high harmonic generation in atoms, molecules and solids (e.g.~\cite{vampa2015linking, vampa2015semiclassical}), 
observation of Bloch oscillations~\cite{schubert2014sub}, 
resolving interfering pathways of electrons in crystals with about 1-fsec precision~\cite{hohenleutner2015real}, inducing~\cite{jimnezgaln2019lightwave} and monitoring topological~\cite{silva2019topological, chacon2020circular, bauer2018high} and Mott insulator-to-metal~\cite{silva2018high} phase transitions, resolving coherent oscillation of electronic charge density order~\cite{nag2019dynamical}, 
identifying the van Hove singularities in the conduction bands~\cite{uzan2020attosecond}, and reconstructing effective potentials
seen by the light-driven electrons with picometer accuracy~\cite{lakhotia2020laser}.

% Indeed, while shaping light pulses on the sub-cycle scale
% is a tool to control the electronic response, the 
% nonlinear electronic response can also be 
% used for characterizing few-cycle light pulses, with 
% the pulse CEP being one of the key parameters. 

Here we apply machine
learning to the analysis of high harmonic generation from a crystal, which allows us to 'kill two birds with one stone': reconstruct
the band structure of the crystal and fully characterize
incident few-cycle laser pulses, including both their
nonlinear chirp and the 
phase of the carrier oscillations under the envelope 
%(the carrier-envelope phase, CEP). 
(CEP).

The fundamental role of the CEP in nonlinear light-matter interaction has been understood theoretically in~\cite{de1998phase, cormier1998effect, tempea1999phase, dietrich2000determining}, 
stimulating first experiments in the gas phase~\cite{paulus2001absolute}. Powerful gas-phase methods for characterizing few-cycle pulses
have been developed,  including stereo-above-threshold ionization (stereo-ATI)~\cite{Zhang:17, paulus2003measurement, milovsevic2006above}, 
attosecond streak camera and its modifications~\cite{Goulielmakis:04, itatani2002attosecond, mairesse2005frequency, baltuvska2003attosecond, kienberger2004atomic}, and half-cycle high harmonic 
cutoffs~\cite{Haworth:07}.
Using nonlinear response of solids for
characterizing the CEP has also been pursued~\cite{Dombi:04, apolonski2004observation, Paasch:13}. 
%and directional currents in nano-junctions 
%\cite{Paasch:13}.
%and dielectrics~\cite{schultze2013controlling}. 
Yet, all-optical, all-solid-state characterization of 
few-cycle pulses, including their CEP,
remains a challenge.
We hope to change this situation.
Particularly relevant to our work
are the earlier proposal  on using interference patterns in spectrally
overlapping regions of even- and odd-order harmonics in solids~\cite{mehendale2000method} and the use of two-color high harmonic spectroscopy for all-optical reconstruction of the band structure~\cite{Vampa:15} from the two-dimensional high harmonic spectra, recorded as a function of the harmonic frequency and the two-color delay.

Two-dimensional spectra
of the nonlinear-optical response may 
provide sufficient information 
to recover the pulse. One prominent example is frequency-resolved optical gating, which
uses the 
second-order optical response 
recorded as a function of the time-delay between the two incident pulses, the target pulses and the auxiliary gate pulse 
(e.g.~\cite{Trebino:93, Miranda:12, mairesse2005frequency}.)
%,
%with the rigorously proven possibility of 
%pulse reconstruction.  
Extending this analysis to 
highly nonlinear optical response remains an open problem. 
The crucial importance of addressing this problem stems from
the fact that such analysis would allow one to characterize
the laser pulse directly in the interaction region.

In special cases, such as the case of
the two-color high harmonic spectroscopy of attosecond 
pulses using fundamental and the second harmonic
(e.g.~\cite{dudovich2006measuring}),
the 2D harmonic spectra recorded as a function of the 
two-color phase and the harmonic frequency
may carry sufficient information for reconstructing 
attosecond pulses as they are produced,
%produced by highly nonlinear optical response 
directly in the  interaction region. Such reconstruction does, however, require detailed understanding of the 
physics of the microscopic quantum response. 
The possibility of solving a full reconstruction 
problem in a general case, recovering both the 
pulse and the quantum system, remains completely 
unexplored. 

% i.e. proving that a functional dependency exists or doesn't exist between the response data and certain underlying parameters we wish to recover, is much requires either an analytical expression for the response function, which is not available for HHG in crystals, or an exact numerical demonstration that a function exists that links the responses to the intrinsic parameters. 

When there is no simple and/or well known functional dependence between 
the response data and the parameters one wishes 
to reconstruct, the problem is well suited for neural networks. Such networks  aim to find a smooth analytical function $f_\theta(x)$ which connects the input $x_i$ and the desired output $y_i$. If it is successful, one can conclude that the data $x$ indeed does contain the information $y$. Pertinent examples include applications to solving the Schr\"{o}dinger equation, where
neural networks can output highly accurate results 
\cite{mills2017deep,giri2021perspectives}.
Our results show that, given sufficient training set, 
the 2D spectra of the 
high harmonic response as a function of the nonlinear response
frequency and the CEP of an unknown driving pulse allow 
for simultaneous reconstruction of both the
pulse and the unknown crystal band structure, even
in the presence of substantial noise including
CEP jitter.

% Here we apply machine
% learning to the analysis of high harmonic generation from a lattice, which allows us to both reconstruct
% the band structure of the crystal and fully characterizing 
% incident few-cycle laser pulses, including both 
% its nonlinear chirp and the 
% phase of the carrier oscillations under the envelope 
% (the carrier-envelope phase, CEP). 

% When this is not the case, such as when dealing with the analytically-intractable HHG in solids~\cite{Vampa:15}, 

\begin{figure}
\begin{center}
    \includegraphics[width=\textwidth]{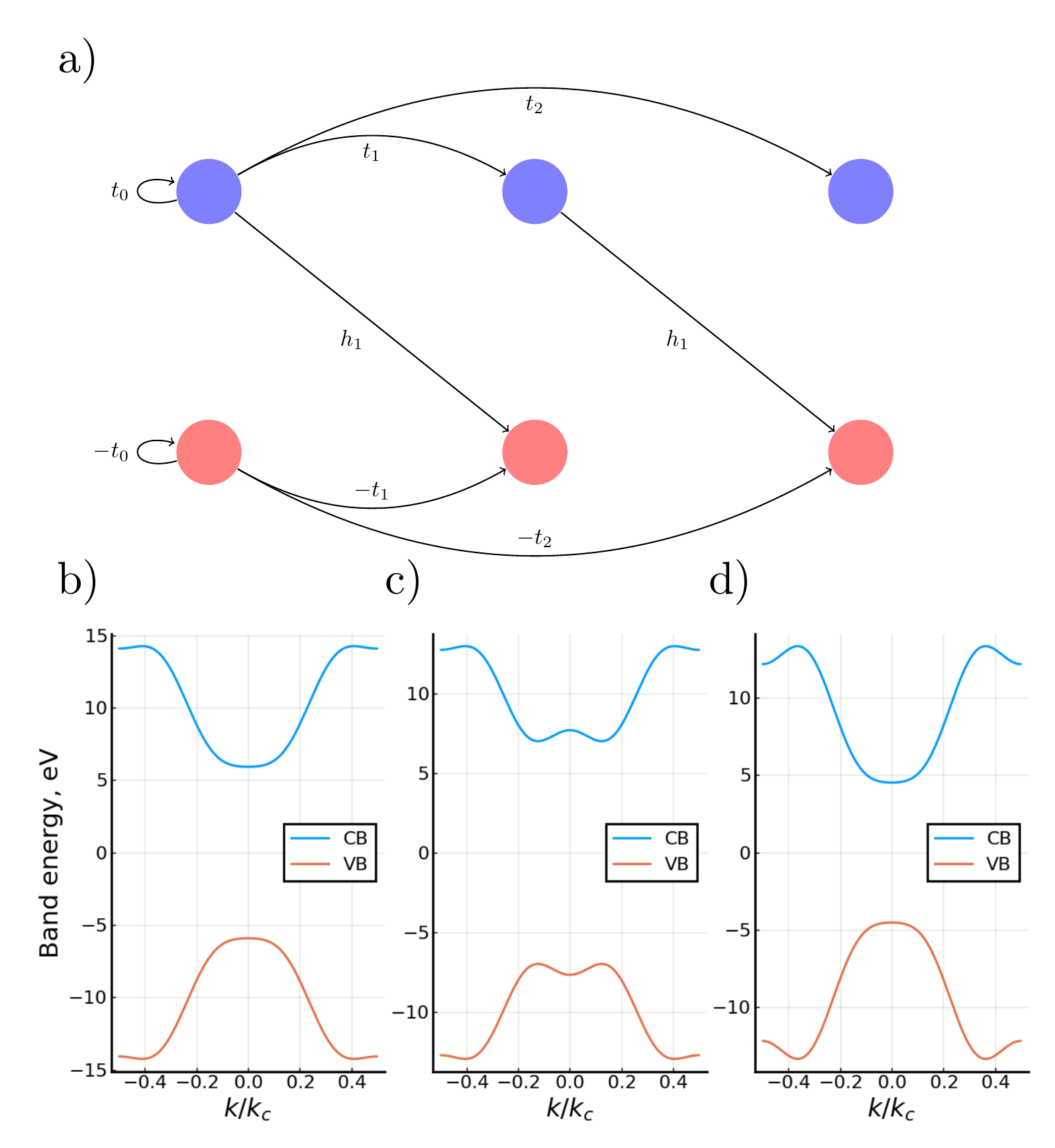}
    \caption{Model system used for pulse reconstruction 
    and band structure spectroscopy via nonlinear-optical response. (a) The two types of sites present, A and B, are connected by hopping constants $t_j$ between similar sites and $h_1$ between the next-neighbor sites of the different
    type.
    (b-d) Examples of band structures that can be generated with this model system.}
    \label{fig:system}
\end{center}
\end{figure}

To demonstrate the method, we assume no apriori knowledge 
about the incident pulse and use rather limited knowledge
about the nonlinear medium. 
For the 
quantum system, we begin with a modified Rice-Mele model~\cite{rice1982elementary} 
with nearest neighbor, next nearest neighbor, etc. hoppings, see Figure \ref{fig:system}(a) (and 
Supplementary information for further details.)
Both the on-site energies and the couplings are 
assumed to be unknown. The reconstruction procedure is expected to output both the parameters of the pulse and of the lattice. We are thus faced with a nonlinear optimization problem with a very large input dimension, which requires finding optimal interpolation between the existing trial samples.

Such a regression tool is provided by deep neural networks (DNNs)~\cite{Carleo:19}, already used for such diverse applications as boosting the signal-to-noise ratio in LHC collision data~\cite{Baldi:14}, establishing a fast mapping between galaxy and dark matter distribution~\cite{Zhang:19}, and constructing efficient representations of many-body quantum states~\cite{Sharir:20}. The inherent resilience of neural networks to noise is an important asset for pulse shape characterization. The emergence of photonic implementations of feed-forward neural networks~\cite{Shen:17} outlines a perspective of implementing this regression scheme in an all-optical way.

%The fact that the magnitude of the 
%intraband hoppings' does not differ for atoms A and B does not lead to %any loss in generality. 

The vector potential of the incident laser field, $A(t)$, is 
generated in the frequency domain with the unknown to the neural network quadratic and the cubic phases:
\begin{equation}
	A(\omega) \propto \exp(i\phi(\omega)) = \exp\left(i\lambda(\omega-\omega_0)^3/6 - \mu(\omega-\omega_0)^2/2 + i\varphi\right)
\end{equation}
The complex parameter $\mu$ is defined 
in such a way that, in the 
absence of the cubic chirp, 
the pulse has a temporal width $\sigma\equiv T_0$: $\mu \equiv \cfrac{\sigma^2 - i\alpha}{1+\alpha^2/\sigma^4}$, where $\alpha$ is the chirp parameter. The chirp parameters $\alpha$ and $\lambda$ are expressed
via dimensionless quantities $\beta$ and $\epsilon$,  
$\alpha = \pi(\sigma/2)^2 \times \beta$, $\lambda = \pi\sigma^3 \times \epsilon$. The dimensionless parameters vary in the ranges $\beta\in\left[-2.0, 2.0\right], \epsilon\in\left[-1.0, 1.0\right]$. Finally, $\varphi$ sets the CEP of the pulse, also unknown to the neural
network and chosen randomly. The system evolution is simulated for 40 laser cycles; an additional Gaussian 
cutoff is introduced at 
the leading and trailing edges of the pulse 
for numerical stability (see Supplementary information.)

% In the time domain, the resulting pulse shape can be written
% in closed analytical form:
% \begin{equation}
% 	A(t) = \cfrac{F_0}{\omega_0}\text{ Im}\left(\mathcal{E}(t) \exp(i(\omega_0 t+\varphi))\right)
% \end{equation}
% The envelope function in the above equation is:
% \begin{equation}
%     \label{eqn:pulset}
% 	\mathcal{E}(t) = 2\sqrt{\frac{\pi|\mu|}{2}} \left|\cfrac{2}{\lambda}\right|^{1/3} \text{Ai}\left(\left(\cfrac{2}{\lambda}\right)^{1/3}\left(t+\frac{\mu^2}{2\lambda}\right)\right)\exp\left(\cfrac{\mu}{\lambda}\left(\cfrac{\mu^2}{3\lambda} + t\right)\right)
% \end{equation}
% The normalizing factor is introduced to keep the peak absolute value of the envelope function at 1 when there's no cubic phase, and conserve the total pulse power at all parameter values. For numerical stability, we introduce a cutoff function that ensures that $A(t)=0$ at the beginning and end of the pulse:
% \begin{equation}
% 	\text{cutf}(t) = \begin{cases} 1.0, & \mbox{if } t>2T \\ \exp\left(-\cfrac{(t-2T)^2}{2(T/2)^2}\right), & \mbox{if } t\leq 2T \end{cases}
% \end{equation}
% with the numerical vector potential taking the form
% \begin{equation}
%     A_{num}(t) = A(t)*\text{cutf}(t)*\text{cutf}(T_{max} - t)
% \end{equation}

\begin{figure}
\begin{center}
    \begin{tikzpicture}%[on grid, node distance=3cm and \textwidth/6]
        \node[inner sep=0] (pic1) at (0, 0) {\includegraphics[width=\textwidth]{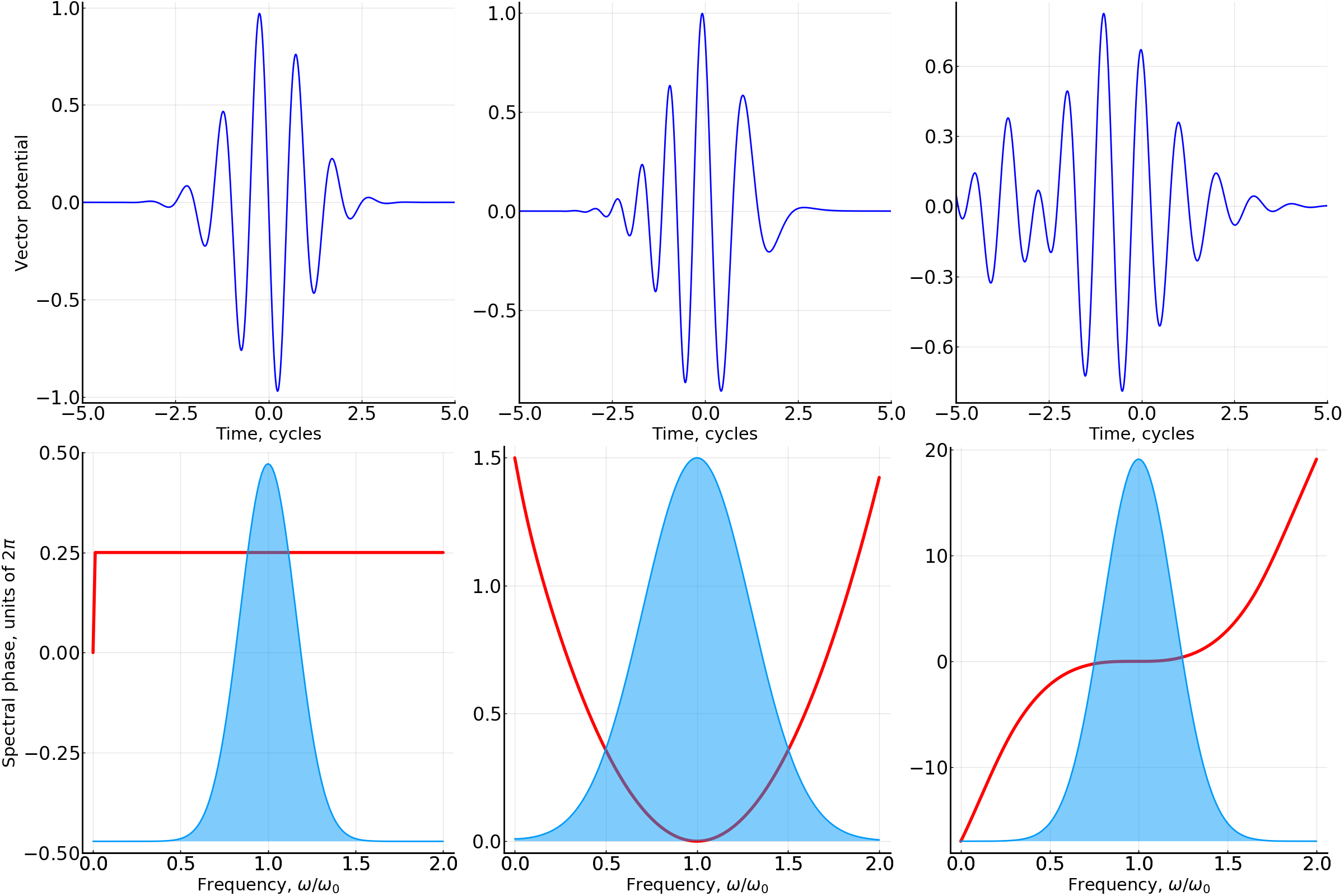}};
        \node[scale=2.0, xshift=0.7cm, yshift=-0.25cm] (ta) at (pic1.north west) {a)};
        \node[scale=2.0,xshift=\textwidth/6] (tb) at (ta) {b)};
        \node[scale=2.0,xshift=\textwidth/6] (tc) at (tb) {c)};
        \path let \p1=(pic1.north),\p2=(pic1.south), \n1={\y1-\y2} in node[scale=2.0, yshift=-\n1/4] (td) at (ta) {d)};
        \node[scale=2.0,xshift=\textwidth/6] (te) at (td) {e)};
        \node[scale=2.0,xshift=\textwidth/6] (tf) at (te) {f)};
    \end{tikzpicture}
    \caption{Sample pulse shapes used in the problem. (a-c) show the time-domain. (d-f) are the frequency-domain
    representations of their respective pulses, where the spectral phase (red line) and spectral amplitude normalized to the central frequency amplitude (blue area), are plotted with respect to frequency. The pulse parameters are $(\varphi, \beta, \epsilon)$, left to right: $(\pi/2, 0.0, 0.0), (0.0, 2.0, 0.0), (0.0, 1.0, 1.0)$.}
    \label{fig:pulses}
\end{center}
\end{figure}

The typical pulses we have used for reconstruction are shown in Fig.~\ref{fig:pulses}, both in time and frequency domain. The latter shows the spectral phases  (red curves) alongside the spectral amplitudes (blue curves), as a function of $\omega/\omega_0$, where $\omega_0$ is the carrier.
Our typical simulated "measurement" assumes that one can 
systematically vary the (unknown) CEP. Thus, we perform
calculations by varying  $\varphi + \Delta\varphi$, with
$\Delta\varphi$ spanning the full range  $\Delta\varphi\in\left[0,2\pi\right)$. For each $\Delta\varphi$, we measure the absolute value of the spectral amplitude 
of the laser-induced current
$|j(\omega)|$, which is given by the Fourier transform
of the calculated current $j(t)$. 

The input data is composed of 
the absolute values of the integer harmonic amplitudes $|j(N\omega, \varphi+\Delta\varphi)|$. The resulting 2D map as a function of $\Delta\varphi$ and $\omega$ is used as the input into the neural network. 
The network must then infer the (randomly chosen in each trial) 
intraband hoppings $\{t_j\}$, the unknown initial CEP $\varphi$, and the pulse parameters 
$\alpha$ and $\lambda$. The values of $\sigma$, frequency $\omega$, and $h_1$ are kept constant throughout. For more information about the inputs used, see Supplementary information.

\begin{figure}[t]
    \includegraphics[width=\textwidth]{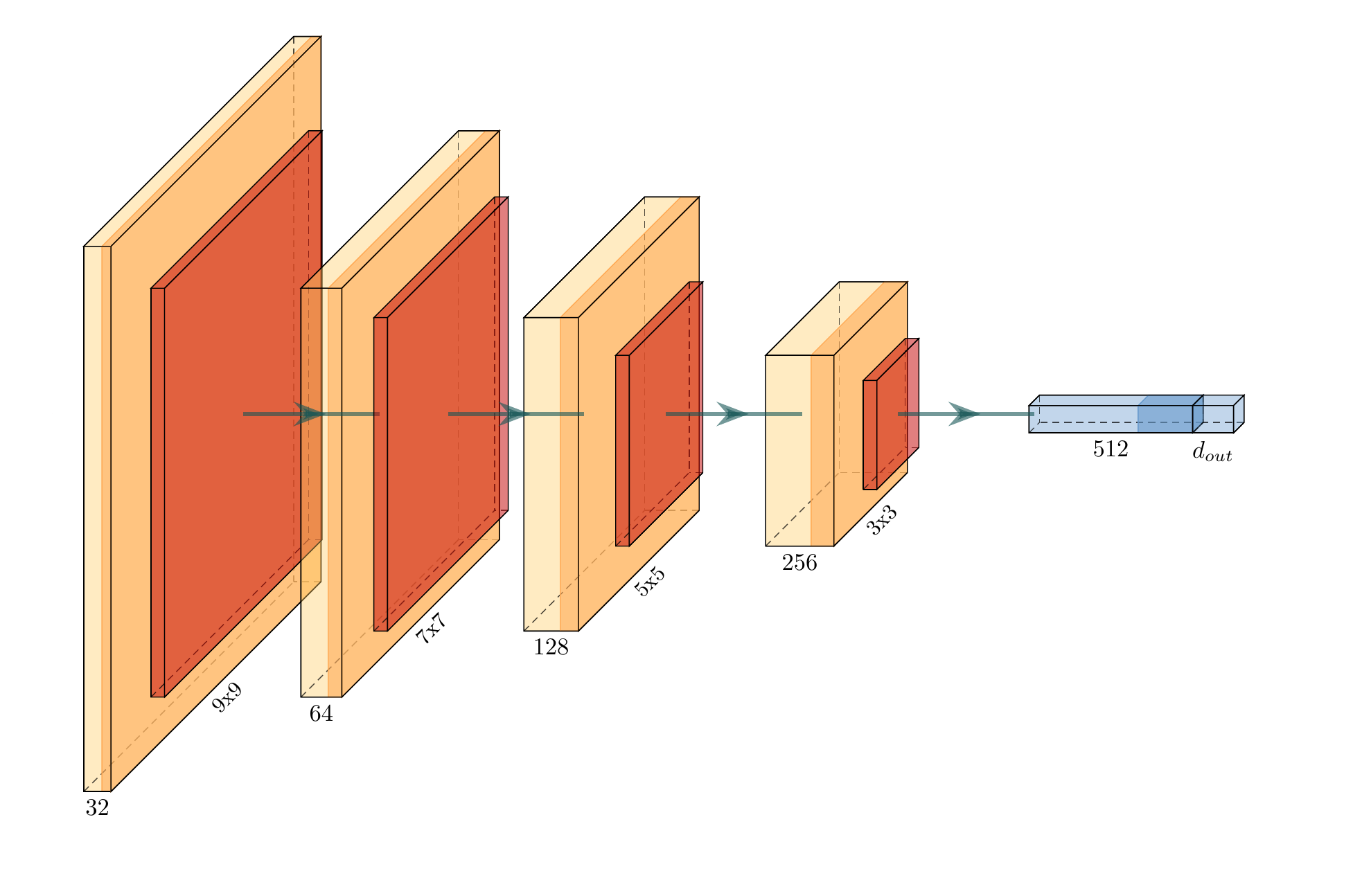}
    \caption{Network architecture used in the paper. Yellow denotes a convolution layer with the number of filters below, orange a batch normalization with subsequent Swish activation, red a maximum pooling layer, blue a fully connected layer, dark blue a Swish activation with no preceding batch normalization. The output of the last FC layer is linear.}
    \label{fig:m-ffnn2}
\end{figure}

In the reconstruction procedure, we have used two separate neural networks, one for recovering the band parameters, and another for recovering the CEP and pulse parameters. Both are constructed using the same architecture, consisting of 4 convolutional layers and two fully-connected ones, see Fig.~\ref{fig:m-ffnn2}. Constructing them as a single neural network with split fully connected layer impedes the overall performance of the scheme. For either problem, this network is trained for 200 epochs with the AdaBelief~\cite{zhuang2020adabelief} optimizer with batch size 256; the learning rate is initially set to $10^{-3}$ and discounted at the 150th epoch by a factor of 10.

To speed up training, we train networks to work on more complex datasets (e.g. with chirped pulses) by using a network pre-trained on a simpler dataset (e.g. with no chirp). The rest of the training procedure remains the same, but the overall number of epochs is reduced to 50.

The recovery results are demonstrated in Fig.~\ref{fig:plots}. The number of unknown band parameters is set to 4. The agreement between the actual and
the reconstructed data, both for the system and for the pulse, is excellent. Tables with detailed analysis of the average performance are given in the supplementary material. 

\tikzstyle{picture}=[inner sep=0, anchor=north]
\begin{figure}
    %\centering
    \begin{tikzpicture}
        %a) \hfill b) \hfill ~\\
        \node[picture] (pic1) at (0, 0) {\includegraphics[width=\textwidth]{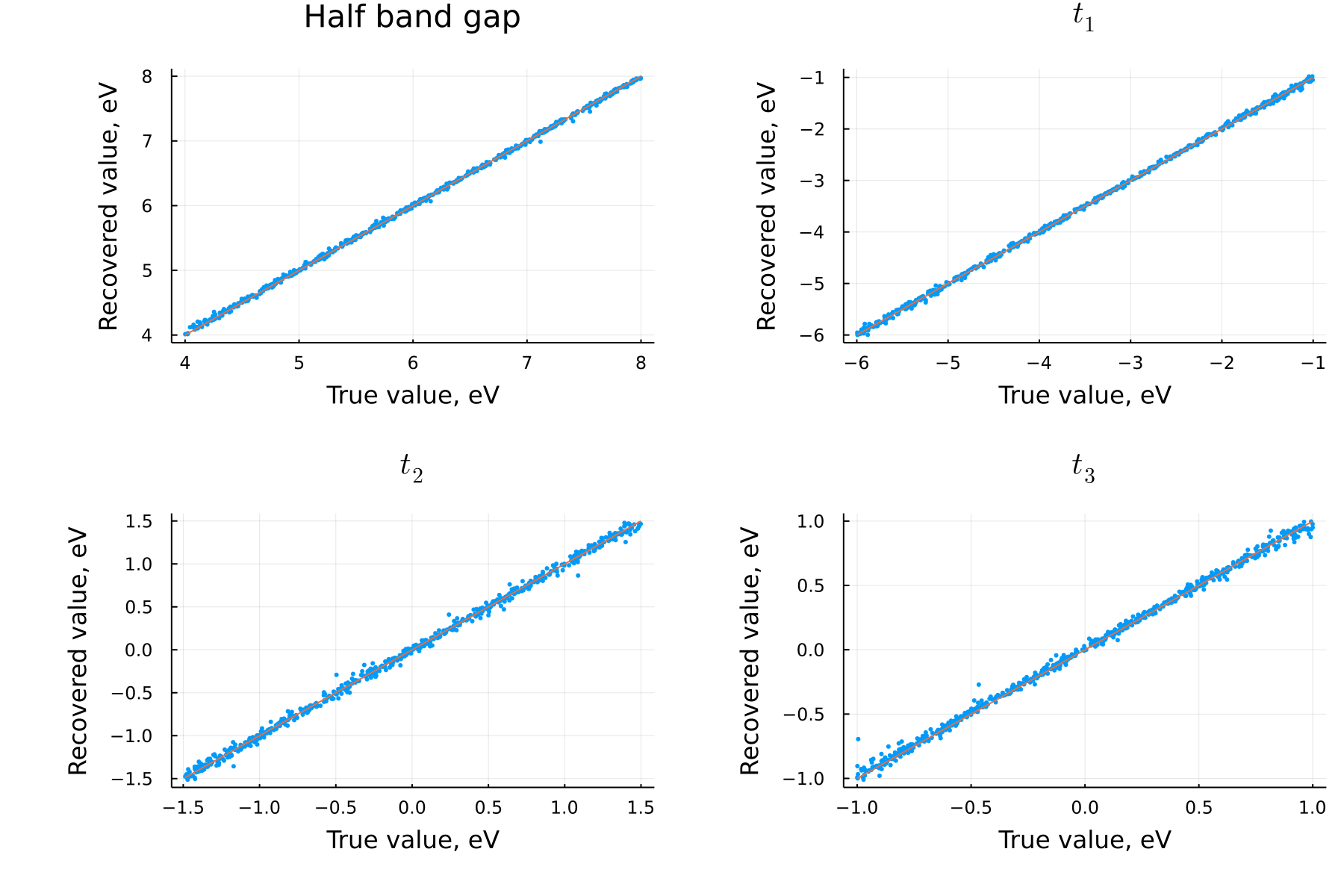}};
        %b)\\
        \node[picture] (pic2) at (pic1.south) {\includegraphics[width=\textwidth]{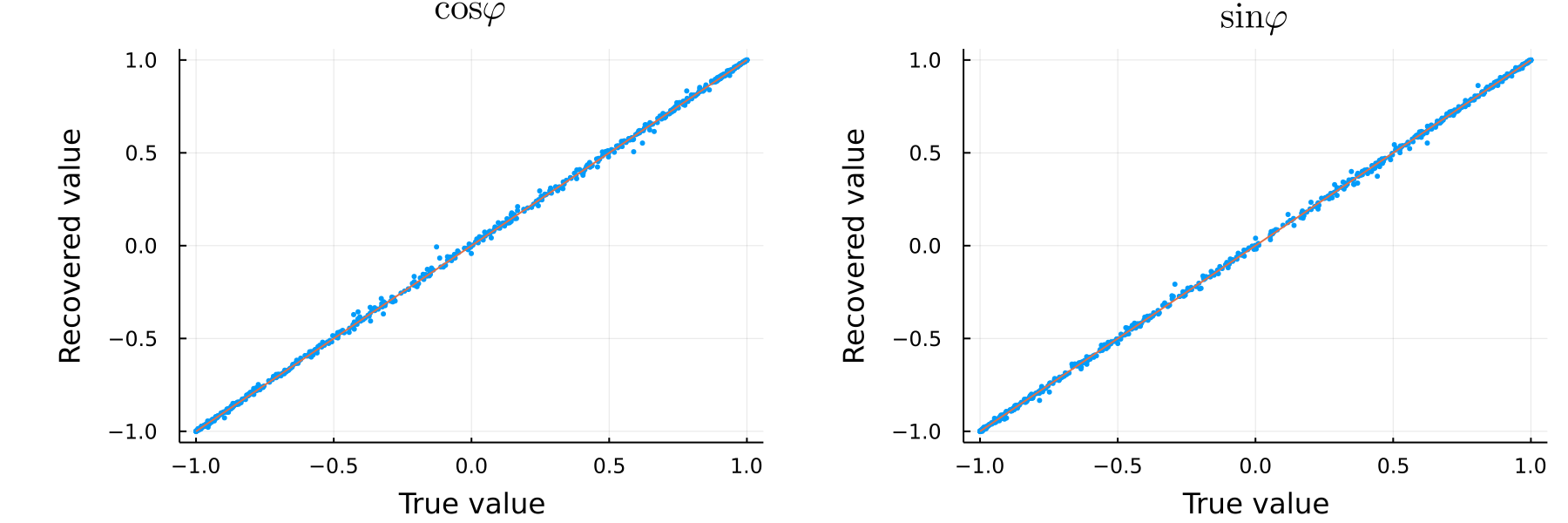}};
        
        \node[scale=2.0, xshift=1.3cm, yshift=-0.4cm] (ta) at (pic1.north west) {a)};
        \node[scale=2.0, xshift=\textwidth/4] (tb) at (ta) {b)};
        \path let \p1=(pic1.north),\p2=(pic1.south), \n1={\y1-\y2} in node[scale=2.0, yshift=-\n1/4] (tc) at (ta) {c)};
        \node[scale=2.0, xshift=\textwidth/4] (td) at (tc) {d)};
        \node[scale=2.0, xshift=1.2cm, yshift=-0.4cm] (te) at (pic2.north west) {e)};
        \node[scale=2.0, xshift=\textwidth/4] (tf) at (te) {f)};
        
        %\node (te) [below=of tc] {e)};
        %\node (tf) [right=of te] {f)};
    \end{tikzpicture}
    \caption{Band parameters (a-d) and CEP (e, f) recovered by the neural network for the source problem (Rice-Mele model, no chirp or cubic phase). The solid orange line $y=x$ serves as a reference. All points used belong to the test set.}
    \label{fig:plots}
\end{figure}

We now also demonstrate that our neural network recovers the system parameters with comparable precision when the input-data contains experiment-like noise. We use a network pretrained on the initial noiseless dataset with no chirp or cubic phase. To simulate experimental noise, we have introduced a random CEP shift to each of the pulses in each sample, sampled from a uniform distribution within $-50\div 50\text{ mrad}$. The output amplitudes were also affected by a uniformly-distributed multiplicative noise with an amplitude of 0.1. The exact performance figures with and without added noise can be found in the supplementary information.

After demonstrating that our neural network recovers the band structure and pulse parameters of the simple source (Rice-Mele) model with high accuracy, we apply our approach to more complex systems. As an example of such more complex system, we used the 2-dimensional gapped graphene system. Its parameters (on-site energy and first-neighbor hopping) were generated in the vicinity of experimental values for hBN, within $4.0\div 7.27$ eV and $0.08\div 0.16$ a.u., respectively.
We simulated its responses by integrating the semiconductor Bloch equations. To simulate experimental conditions, we applied noise using the same procedure as described above.

\begin{figure}[t]
    %\centering
    \resizebox{\textwidth}{!}{
    \begin{tikzpicture}
        %\node[picture] (pic1) at (0, 0) {\includegraphics[width=\textwidth/2]{pictures/accuracy-0605_1337.png}};
        \node[picture] (bands) at (0, 0) {\includegraphics{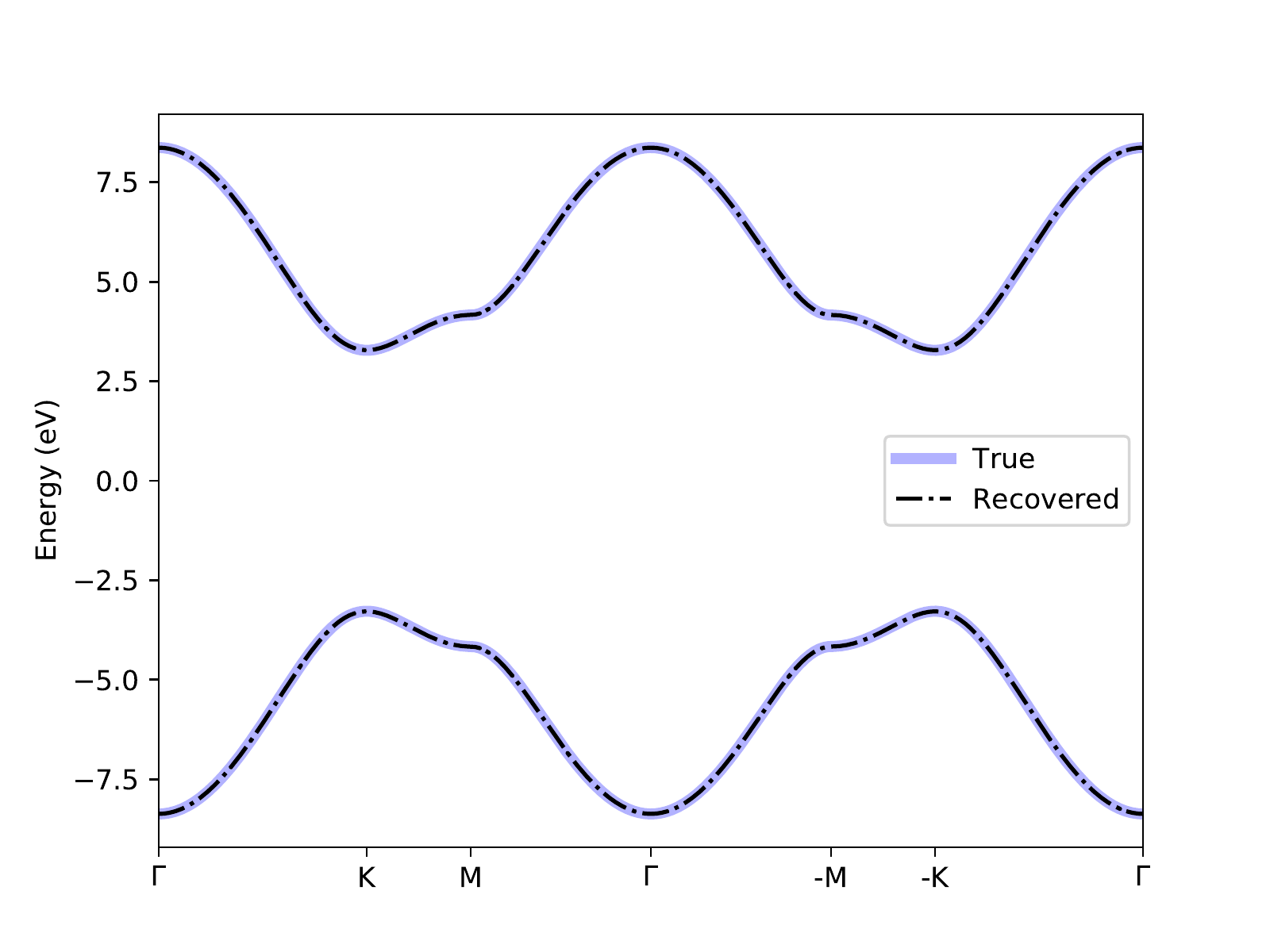}};

        \node[picture, anchor=west] (vpot) at (bands.east) {\includegraphics{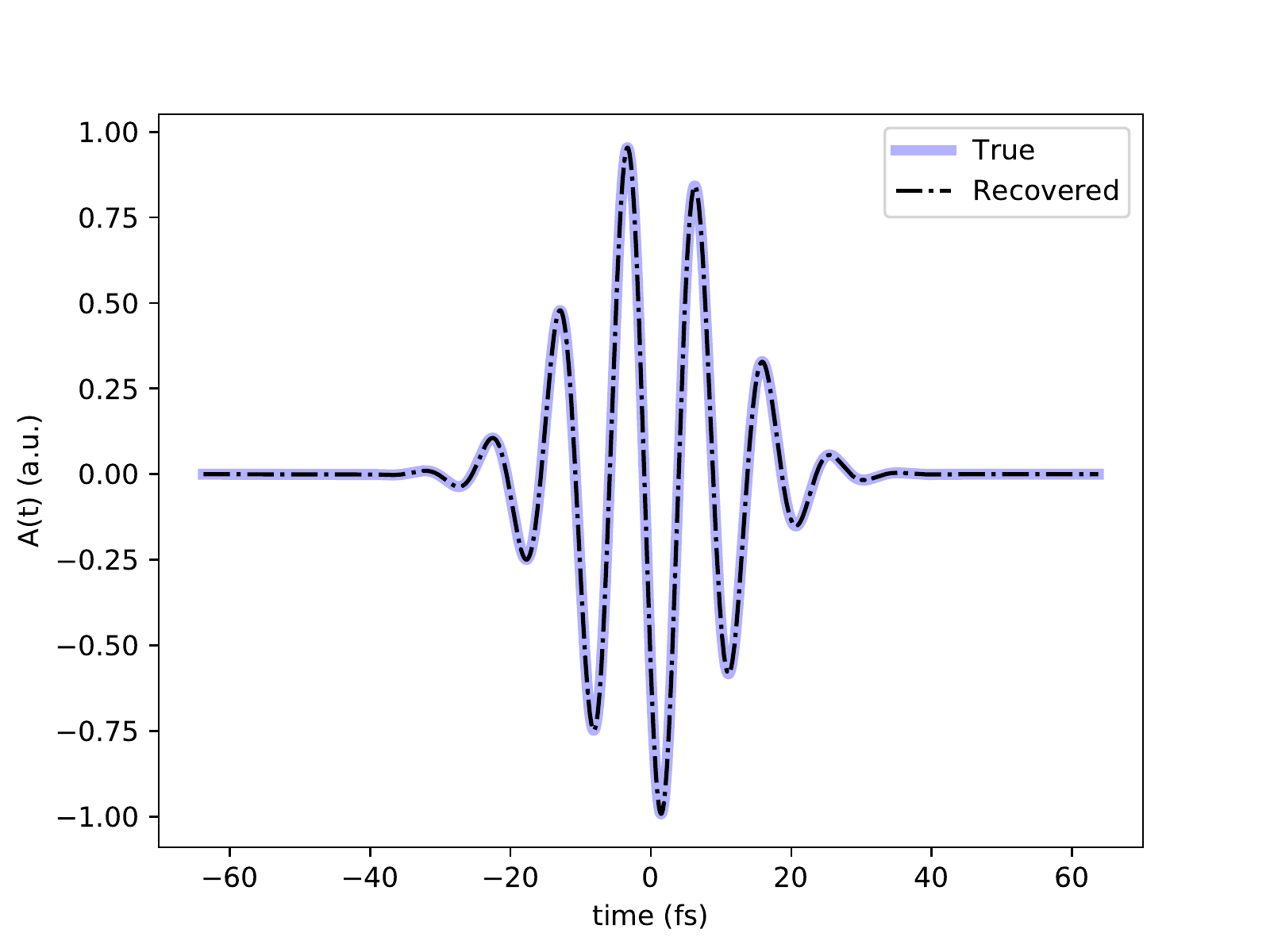}};
        
        \path let \p1=(vpot.east),\p2=(bands.west), \n1={\x1-\x2} in node[picture, anchor=north west, xshift=-1.5cm, text width=\n1] (pic2) at (bands.south west) {\includegraphics[width=\textwidth]{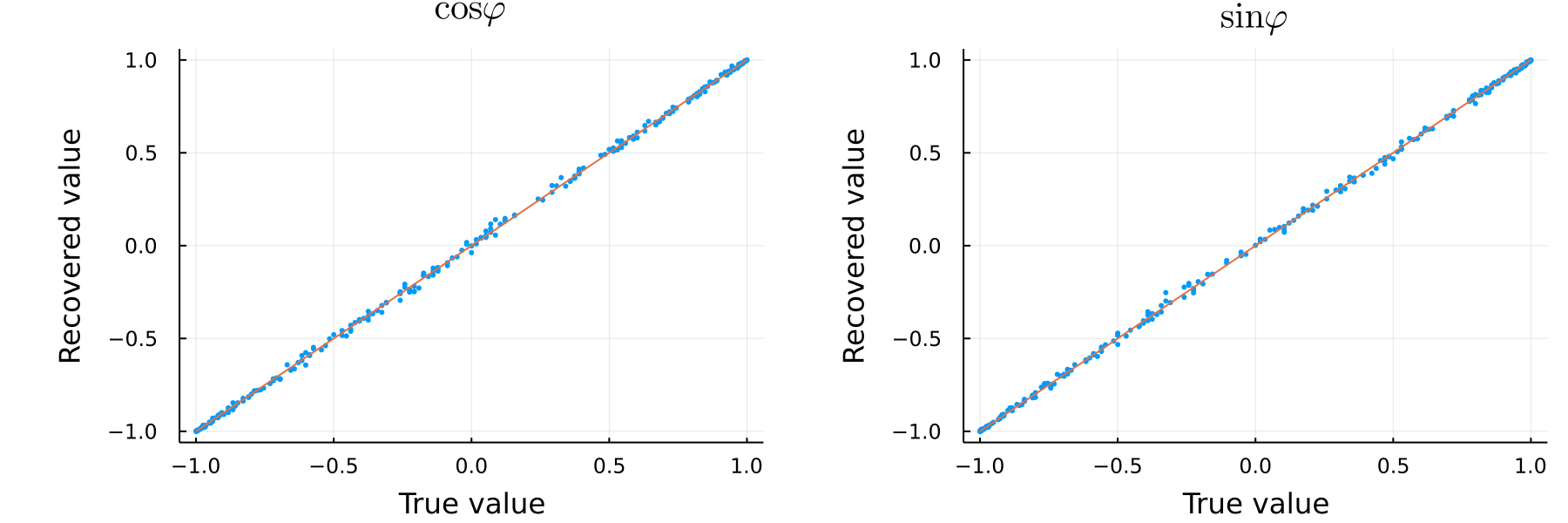}};
        
        \node[scale=4.0, xshift=1.2cm, yshift=-0.7cm] at (bands.north west) {a)};
        \node[scale=4.0, xshift=1.0cm, yshift=-0.7cm] at (vpot.north west) {b)};
        
        %\node[xshift=1.2cm, yshift=-0.6cm] (ta) at (pic1.north west) {a)};
        %\node[xshift=\textwidth/4] (tb) at (ta) {b)};
        \node[scale=4.0, xshift=1.4cm, yshift=-0.6cm] (td) at (pic2.north west) {c)};
        \node[scale=4.0, xshift=1.4cm, yshift=-0.6cm] (te) at (pic2.north) {d)};
        %\node (tb) [right=of ta] {b)};
        
        %\node (tc) [below=of ta] {c)};
    \end{tikzpicture}}
    \caption{Band structure (a), pulse waveform (b),  and CEP (c, d) recovered by the neural network for the target problem (2-dimensional gapped graphene with phase and multiplicative noise).}
    \label{fig:plots-ret}
\end{figure}

Due to the greater computational cost, we could only use 1280 samples as opposed to 65536 for the source problem. 
Such a dataset is too small to train an entire new network. Instead, we applied the transfer learning approach by using the networks pre-trained on datasets with no quadratic or cubic phase, varying the CEP, and 4 unknown band parameters, and partially retraining them (see Supplementary). We discovered that, in spite of the greater complexity of the underlying physical system, such a setup recovers the CEP with a precision comparable to the original problem (see Figs.~\ref{fig:plots-ret}), and achieves good 
relative accuracy on the band parameters. 

This adds new significance to the achieved results. Indeed, 
we have now demonstrated that the spectra from the initial model, while only resembling a real-world setup in a qualitative sense, provide a useful training ground for the neural network, allowing it to acquire useful abstract concepts (parameterized by the deeper layers) which it can later apply to more practical problems, 
for which it was also harder to generate as many 
training samples. 
%Although real-world experimental spectra generated by CEP-stable pulses in solid-state systems with controllably-tuned and monitored CEP are scarce, one can now reasonably expect to apply our approach to experimental data.

Our results demonstrate that solid-state HHG spectra contain more information than one extracts with conventional methods: not only
the pulse parameters, but also the parameters of the quantum system are robustly reconstructed. Our approach replicates the advantages of gas-phase HHG spectroscopy, namely, the ability to resolve the CEP of laser pulses and the complex pulse shapes with polynomial spectral phase nonlinearities. At the same time, 
it requires neither the XUV pulses nor the photoelectron spectroscopy, such as the stereo-ATI. Moreover, it allows for all-optical solid-state implementation. 
In terms of required observables, it is closest to the method based
on measuring the half-cycle cutoffs in gas-phase high harmonic
generation~\cite{Haworth:07}. However, it also allows one to deal with very strong chirps. Applying neural networks to the analysis
of half-cycle cutoffs and high harmonic generation spectra
in the gas phase could be very interesting, especially in molecules, where multiple coupled harmonic generation channels present
challenges for unravelling the underlying laser-driven 
multi-electron dynamics~\cite{bruner2016multidimensional}. 

One could apply the developed 
neural network design to standard gas-phase experiments, 
to analyse the possibility of resolving the 
spectral phase and the CEP for short pulses by processing HHG spectra generated by known inert gases (Ar, Ne, etc.). In this
case the neural network can be trained using TDSE simulations of the necessary responses before being applied to real experimental data.

% The idea of using neural networks to extract information from HHG spectra of analytically intractable systems can be advanced further. Our method could be applied to resolving Floquet levels and bands for crystals and molecules, respectively, which will open unique opportunities for investigating dynamics of complex systems.

The key difficulty of using high harmonic spectroscopy in solids 
is that, without apriori knowledge of the band structure, one lacks 
closed-form solutions for electron dynamics, similar to those available in the gas-phase. Our method circumvents this difficulty. Pulse 
characterization device implementing our principles could be 
tabletop, all solid-state, and capable of operating at ambient conditions.

Another interesting direction to pursue would be to apply novel physics-informed neural network architectures\cite{greydanus2019hamiltonian, toth2019hamiltonian} to resolve Hamiltonians of systems with many degrees of freedom (such as molecules) using 
time-resolved HHG spectra, such as those
obtained from solids driven by mid-IR fields~\cite{hohenleutner2015real}. Neural networks can also
be used for processing the sets of harmonic spectra connected 
by other relations, such as being measured for different angles
between the crystal axes and the driving field, to uncover
effective laser-modified potentials for the charge motion,
extending the pioneering work in Ref.~\cite{lakhotia2020laser} to recover effective potentials of active band electrons. Here, once again, one can take advantage of our idea of using neural networks to extend a method of processing analytically-tractable systems to intractable ones, recovering effective structures.

\section{Funding}
M.I. and N.K. acknowledge funding of the DFG QUTIF grant IV152/6-2. N.K. acknowledges funding by the Foundation for Assistance to Small Innovative Enterprises (agreement No 196GUTsES8-D3/56338) and Foundation for the Advancement of Theoretical Physics and Mathematics (agreement No 20-2-2-39-1). This project has received funding from the European Union’s Horizon 2020 research and innovation programme under grant agreement No 899794.
\section{Acknowledgements}
We thank Vera V. Tiunova for her useful feedback. N.K. and M.I. developed the idea behind the paper and wrote the manuscript. N.K., \'A. J.-G. and R.E.F.S. performed the calculations. N.K. designed the neural network architecture.
\section{Disclosures}
The authors declare no competing interests.

\section{Data availability}
The generated datasets, trained neural network weights, and reconstructed values are available in Dataset 1 (Ref.~\cite{dataset}). The codes running the TDSE simulation and neural network reconstruction of parameters were written in the Julia language~\cite{bezanson2017julia} and are available publicly on GitHub~\cite{codes}.

\bibliography{refs}

\end{document}

\title{Deep neural networks for high harmonic spectroscopy in solids: supplemental document}
\author{}

%\maketitle

\section{Modified Rice-Mele model}

The system is described by the Hamiltonian
\begin{equation}
    \hat{H} = \left[\frac{1}{2}\sum\limits_{j=0}^N t_j \sum\limits_{\alpha\in\{A, B\}} s_\alpha \ket{i\alpha} \bra{i+j,\alpha} + h_1 \sum\limits_{i=-\infty}^{+\infty} \ket{i, A} \bra{i+1, B}\right] + \text{h.c. }
\end{equation}
Here $s_\alpha = -1\text{ for }\alpha=A$ and $s_\alpha = +1\text{ for }\alpha=B$, $N$ is the maximum hopping order, which was varied from 2 to 6. The onsite energies are thus
$-t_0$ for site A and $+t_0$ for site B, while  the 
hopping terms $t_{j\geq 1}$ connect the sites of the same kind. In the simulations, $t_j$ vary randomly between 
training samples and are unknown to the neural network. 
These hopping terms are selected in such a way that the band gap always lies between $8.0$ and $16.0$ eV. The higher hoppings $\{t_1, \ldots\}$ are randomly generated within the ranges: $\left[-6.0, -1.0\right]$, $\left[-1.5, 1.5\right]$, $\left[-1.0, 1.0\right]$, respectively. In all our computations, we assume $\hbar=a=1$, where $a$ is the lattice constant.

The Hamiltonian also includes a term proportional to
$h_1$, which describes hopping between sites $A$ and $B$. 
This term explicitly breaks the inversion symmetry of the model, 
so that its nonlinear-optical 
response allows us to distinguish between pulses with CEPs 
differing by $\pi$.
We set $h_1$ to be much smaller than the typical values of
$t_j$; we used  $h_1 = 0.01\text{ eV}$ and 
$h_1 = 0.05\text{ eV}$. This simplifies modelling relaxation. 
This system allows one to generate a rich variety of band structures and
band-gaps $\epsilon_g(k)$, see Fig.~1 (b-d) for
some examples. 

When we set the lattice constant $a=1, k_c=2\pi$, in the $k$ space, the field-free Hamiltonian transforms to:
\begin{equation}
    \label{eqn:hamiltonian}
    \hat{H}(k) = \epsilon(k)\sigma_z + h_1 \left(\sigma_x\cos k  + \sigma_y\sin k \right)
\end{equation}
\begin{equation}
    \epsilon(k) = \sum\limits_{j=0}^N t_j \cos(jk)
\end{equation}

The field is introduced via the Peierls substitution, which 
transforms the laser-driven Hamiltonian in the 
momentum space as:
\begin{equation}
    \hat{H} = \int dk \ket{k}\hat{H}(k+A(t))\bra{k}
\end{equation}

The system is thus decomposed into an ensemble of non-interacting two-level systems, characterized by the canonical momentum $k$. We evolve each TLS using the master equation with a decoherence term:
\begin{equation}
    \label{eqn:master}
    \dot{\rho}(k) = i[\rho, H(k+A(t))] - \rho_{od}(k)/T_2
\end{equation}
Here $\rho_{od}$ denotes the off-diagonal part of the density matrix in the Hamiltonian eigenstate basis, $T_2$ is the dephasing time, set to $T_0/2$, where $T_0$ is a single fundamental laser period.

\section{Driving pulses in time domain}

In time domain, the pulse shape 
corresponding to the frequency-domain 
representation in the main text can be written
in closed analytical form:
\begin{equation}
	A(t) = \cfrac{F_0}{\omega_0}\text{ Im}\left(\mathcal{E}(t) \exp(i(\omega_0 t+\varphi))\right)
\end{equation}
The envelope function in the above equation is:
\begin{equation}
    \label{eqn:pulset}
	\mathcal{E}(t) = 2\sqrt{\frac{\pi|\mu|}{2}} \left|\cfrac{2}{\lambda}\right|^{1/3} \text{Ai}\left(\left(\cfrac{2}{\lambda}\right)^{1/3}\left(t+\frac{\mu^2}{2\lambda}\right)\right)\exp\left(\cfrac{\mu}{\lambda}\left(\cfrac{\mu^2}{3\lambda} + t\right)\right)
\end{equation}
The normalizing factor is introduced to keep the peak absolute value of the envelope function at 1 when there's no cubic phase, and conserve the total pulse power at all parameter values. For numerical stability, we introduce a cutoff function that ensures that $A(t)=0$ at the beginning and end of the pulse:
\begin{equation}
	\text{cutf}(t) = \begin{cases} 1.0, & \mbox{if } t>2T \\ \exp\left(-\cfrac{(t-2T)^2}{2(T/2)^2}\right), & \mbox{if } t\leq 2T \end{cases}
\end{equation}
with the numerical vector potential taking the form
\begin{equation}
    A_{num}(t) = A(t)*\text{cutf}(t)*\text{cutf}(T_{max} - t)
\end{equation}

\section{Evaluation of the Airy function}

We have approximated the Airy function required for computing the pulse shape in \eqref{eqn:pulset} with Maclaurin series~\cite[Eq.~9.4.1]{NIST:DLMF} and asymptotic series~\cite[Eq.~9.7.5, 9.7.9]{NIST:DLMF} for small and large arguments, respectively. For $|z|<3.5$ we used the Maclaurin series up to $N=25$; for $|z|\geq 3.5$, the asymptotic expansion for $N=5$ was used. This approximation allows for a maximum relative error below $\delta = 3\cdot 10^{-4}$. 

\section{Solution of the Time dependent Schr\"{o}dinger Equation.}

We discretize the 1D Brillouin zone of the crystal into 32 $k$ points (finer discretization does not worsen the NN's performance). For each point, we initialize the system in the pure ground state of the Hamiltonian \eqref{eqn:hamiltonian} for $A(t_0) = 0$, then evolve this state according to \eqref{eqn:master}. The dephasing time $\tau_2 = T_0/2$ is selected phenomenologically. For all computations, the field $F_0=4.0\text{ eV}$, and the central frequency $\omega_0=0.8\text{ eV}/\hbar$ are kept constant.

80 times per cycle, we record the current, computed as:

\begin{equation}
    \hat{\mathcal{J}}(k) = \cfrac{\partial \hat{H}(k)}{\partial k} = -\sum\limits_j j t_j \sin(jk) + h_1(-\sigma_x \sin(k) + \sigma_y \cos(k))
\end{equation}
\begin{equation}
    \braket{j}_k(t) = \text{tr}(\rho_k(t)\mathcal{J}(k+A(t)))
\end{equation}
\begin{equation}
    \braket{j}(t) = \sum\limits_k \braket{j}_k(t)
\end{equation}
In the above equation, we neglect the $2\pi/N_k$ ($N_k=32$) factor. \\
Each "recording" interval is separated into 2 integration steps. Each integration step consists of $\nu=4$ iterations of the following procedure:
\begin{equation}
    \rho \rightarrow \exp(-i H(k+A(t))dt/\nu) \rho \exp(i H(k+A(t))dt/\nu)
\end{equation}
\begin{equation}
    \rho \rightarrow \rho - (dt/\nu)*\rho_{od}/T_2
\end{equation}

Within each integration step, the field $A(t)$ is kept constant. The matrix exponent is computed numerically exactly in the assumption that $H(k)$ is traceless. 

%\section*{Supplementary information}

\section{Neural networks}
Here we first give a very brief general introduction into neural networks, and then discuss applications to our particular problem.

Suppose we search for a model that best describes a known experimental data set $(\mathbf{x}_i, \mathbf{y}_i)$ with $K$ total samples. Generally, the workflow is:
%\newpage
%[topsep=0pt, partopsep=0pt, itemsep=2pt,parsep=2pt]

\begin{enumerate}
    \item Define the model as a trial function $f_\theta(\mathbf{x})$ that depends on a vector of parameters $\boldsymbol{\theta}$.
    \item Define a measure of divergence between the results predicted by the model $\{f_\theta(\mathbf{x}_i)\}$ and experimental results $\{\mathbf{y}_i\}$, expressed as $D = \cfrac{1}{K}\sum\limits_{i=1}^K\mathcal{L}(f_\theta(\mathbf{x}_i), \mathbf{y}_i)$. The most common measure for $D$ is the mean squared error (MSE), obtained for $\mathcal{L} = ||f_\theta(\mathbf{x}_i) - \mathbf{y}_i||^2$.
    \item Optimize the parameters $\boldsymbol{\theta}$ to minimize the divergence $D$.
\end{enumerate}

The most basic example of such an approach is the linear regression: we define the model as $f(\mathbf{x}) = \hat{W}\mathbf{x} + \mathbf{b}$ with $\theta \equiv \{\hat W, \mathbf b\}$, and use the MSE divergence. Minimizing the divergence with respect to $\hat{W}$ and $\mathbf{b}$, we obtain what's known as the linear least squares method. For this case, we have two options available. First, we could find the optimal $\mathbf{\theta}$ analytically by and solving the overdefined system of linear equations.%~\footnote{\url{https://en.wikipedia.org/wiki/Ordinary_least_squares#Matrix/vector_formulation}}%
However, this may not be optimal for high-dimensional data, and so we can choose an alternative way where we optimize the same parameters with gradient descent, updating $\boldsymbol{\theta}$ against the gradient of the divergence function: $\delta\theta \propto -\nabla_\theta \sum_i\mathcal{L}(f_\theta(\mathbf{x}_i), \mathbf y_i)$. Obviously, the result must be the same, which can be verified analytically.

Evidently, linear regression cannot model most data sets. The simple nonlinear regression methods generally rely on fitting the experimental data to an ansatz selected by hand. Such methods include logistical regression, polynomial regression, and the more exotic expansions over various bases such as the Legendre polynomials and harmonic functions. However, selecting the correct nonlinear regression requires extensive trial-and-error testing of various ans\"atze, which is generally not possible except for simple cases with known symmetries and properties of the experimental data set. A convenient solution to this complication is using neural networks, which serves as a universal ansatz by being capable of approximating any continuous finite function to an arbitrary precision~\cite{pinkus1999approximation}. This property is the key advantage of neural networks, which allows them to be successfully applied even to data sets where the very existence of a connection between the input and the output is not evident.

\begin{figure}[H]
    \centering
    \includegraphics[width=\textwidth/2]{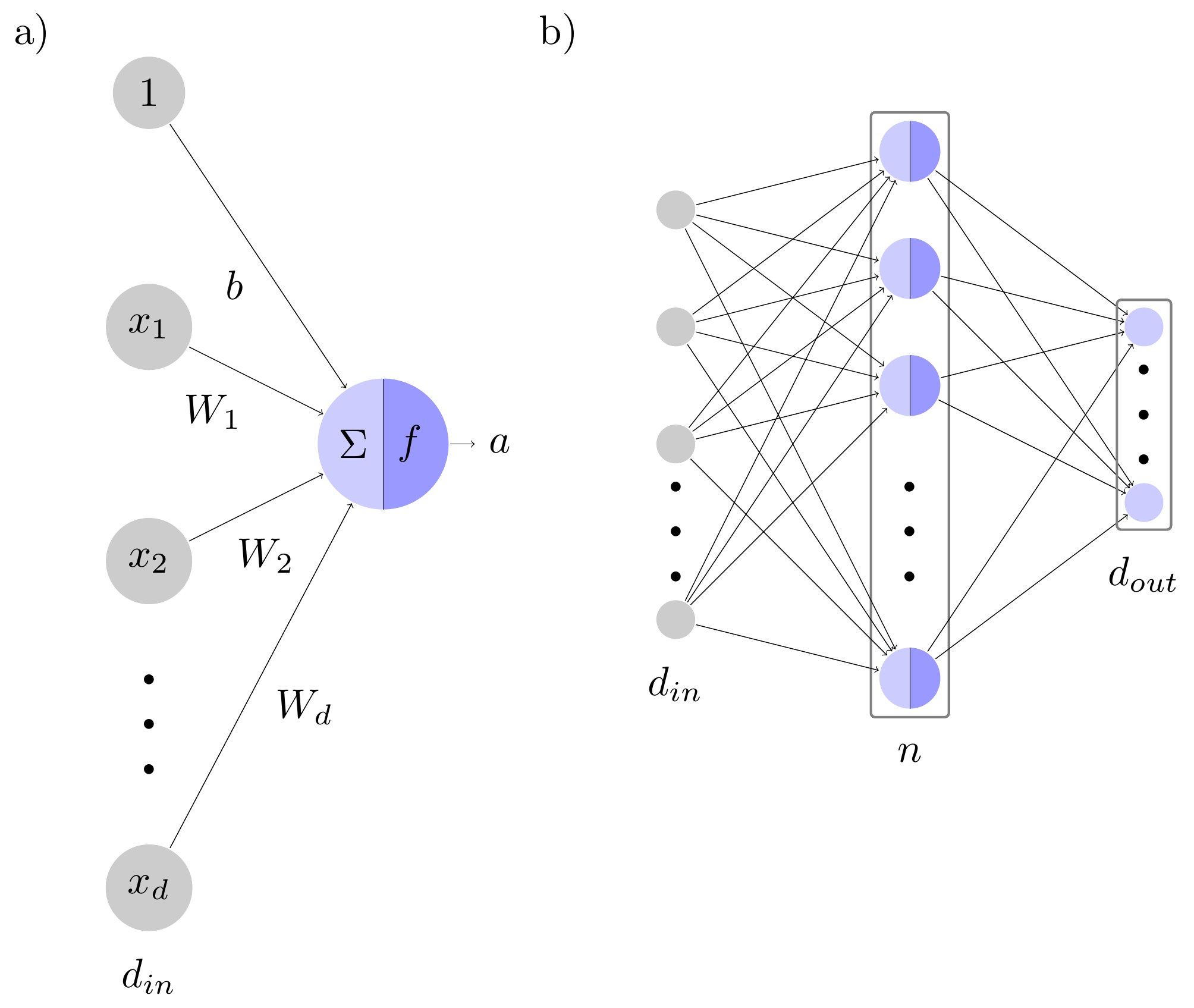}
    \caption{(a) The vector $\mathbf{x}$ is the $d$-dimensional input vector, $\mathbf{W}$ is the $d$-dimensional weight vector, $b$ is the bias. The output is computed as $a = f(\mathbf{Wx}+b)$, commonly rewritten as $a = f(\mathbf{W}^{[1]}\mathbf{x}^{[1]})$ with $W^{[1]}_0=b$, $x^{[1]}_0=1$. The gray circles represent elements of the input data, light blue stands for the weighted summation (linear transform), and dark blue for the nonlinear activation function.
    (b) The layout of a perceptron with a single hidden layer.}
    \label{fig:neuron}
\end{figure}

We will now discuss the core principles of a neural network. A neural network is constructed of building blocks called neurons. The neuron, demonstrated schematically on Fig.~\ref{fig:neuron}(a) is defined as follows.

The neuron is a function with $n$-dimensional input $\mathbf{x}$ and one-dimensional output $a \equiv f(\mathbf{Wx}+b)$. Here $\mathbf{W}$ is the weight vector of the neuron, $b$ is the bias, and $f$ is the activation function. The typically chosen activation functions include the logistical function, ReLU, Softplus, Swish, etc. A neuron encodes a single nonlinearity, such as a smooth step for the logistical ($\equiv$ sigmoid) activation function. Neural networks use compositions of these nonlinearities to approximate arbitrary smooth functions.

In a neural network, the neurons are arranged in layers where each neuron has the same input vector $\mathbf{x}$ and the same activation function $f$. For this layer, the bias $\mathbf{b}$ is expressed by a vector, the weights $\hat{W}$ by a matrix, and the output is given by $\mathbf{a} = f(\hat{W}\mathbf{x}+\mathbf{b})$, where $f$ is applied individually to each argument. The simplest neural network, called the single-layer perceptron, consists of only two such layers, and is schematically demonstrated on Fig.~\ref{fig:neuron}(b). The first (hidden) layer contains all the nonlinearities, and gets the raw data as the input. The second (output) layer may or may not be linear, i.e. have an identity activation function, and in the linear case simply applies a learnable linear transform to the outputs of the hidden layer.

The universal approximation theorem~\cite{pinkus1999approximation} states that any continuous finite function can be approximated by a perceptron with a sufficiently large hidden layer with a non-polynomial activation function. However, for many cases this turns out to be computationally non-optimal. A more efficient representation that requires less parameters can be achieved by using stacked layers, with the output of $k$-th layer being the input for $k+1$-th. A network constructed in such a way is commonly named "fully connected". For N-layer deep fully connected networks, the equation $\mathbf{a} = f(\hat{W}\mathbf{x}+\mathbf{b})$ is commonly rewritten as $\mathbf{a}^{[k]} = f^{[k]}(\hat{W}^{[k]}\mathbf{x}^{[k]})$, where $k$ is the layer number, $W^{[k]}_{i0} = b_i$, $x^{[k]}_0 = 1$. The output function is therefore given by: $f_\theta(\mathbf{x}) = f^{[N]}(\hat{W}^{[N]}\mathbf{x}^{[N]}) = f^{[N]}(\hat{W}^{[N]}f^{[N-1]}(\hat{W}^{[N-1]}f^{[N-2]}(\ldots)))$.

\begin{figure}[H]
    \centering
    \includegraphics[width=\textwidth/2]{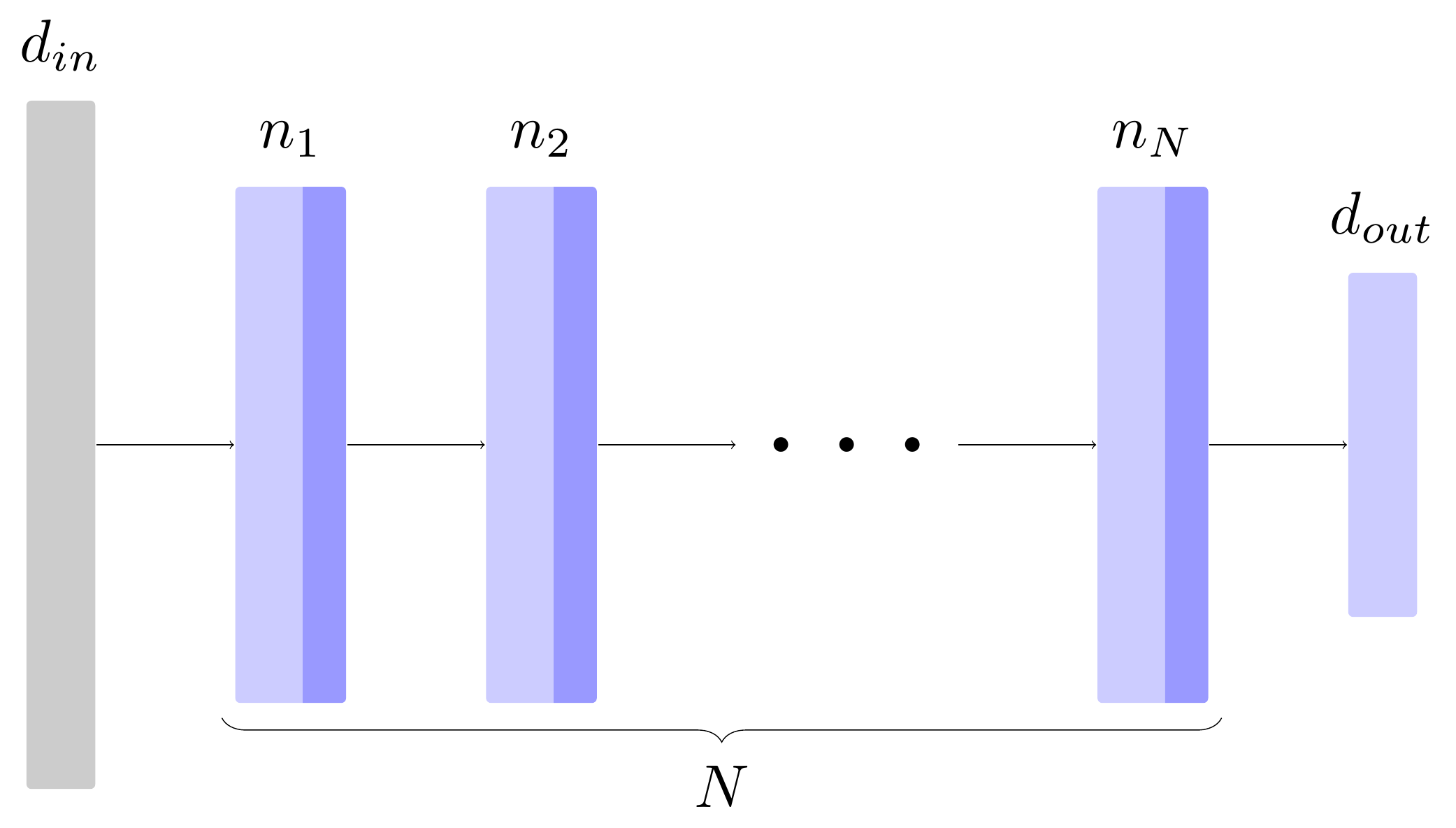}
    \caption{In a deep feed-forward neural network, the output of each layer is used as the input of the next one, with the network output being evaluated at the last layer.}
    \label{fig:ffnn}
\end{figure}

The general heuristic states that the higher layers process higher-level features, and thus can learn highly nonlinear and nonlocal patterns. A 'dual' version of the universal approximation theorem~\cite{lu2017expressive, kidger2020universal} demonstrates that an arbitrary continuous function can also be approximated by a neural network with a fixed (but not arbitrarily small) number of neurons per layer and sufficiently large number of layers. Shown on figure \ref{fig:ffnn} is the general scheme of a deep fully connected neural network.

Deep neural networks can be trained in an optimal way thanks to a procedure called 'backpropagation'~\cite{rumelhart1986learning} that allows to efficiently differentiate the loss function with respect to each of its parameters $\theta \equiv \{\hat W^{[N]}\ldots \hat W^{[1]}\}$. This enables optimizing them with the gradient descent procedure, which consists of updating $\theta$ against the gradient of the divergence function, called 'loss' for neural networks: $\delta\theta \propto -\nabla_\theta \sum_i\mathcal{L}(f_\theta(\mathbf x_i), \mathbf y_i)$, ideally yielding a function that minimizes the error for each sample simultaneously. However, the 'naive' gradient descent is vulnerable to noise and local minima, and in practice, more sophisticated methods are employed, such as ADAM~\cite{kingma2014adam}. ADAM is based on correcting the gradient descent update by using an adapted momentum of the parameters which prevents them from getting stuck in a local minimum. For the same goal, the input data is separated into 'batches' that consist of a defined number of samples. The parameters are updated after processing each batch. The process of optimizing the FFNN parameters, or 'training' the FFNN, therefore consists of the following steps.
\begin{enumerate}
    \item Select a batch of data as input. 
    \item (forward pass) Evaluate the output of each consecutive layer to obtain the NN output.
    \item Compute the loss function between the produced output and the desired output.
    \item (backward pass) Calculate the gradient of the loss function with respect to each parameter by reverse propagating the loss gradient from the last to the first layer.
    \item Compute the update to the parameters using a parameter update procedure (gradient descent, ADAM, ADAGrad, ADAMax, etc) and apply it.
\end{enumerate}
The above steps are repeated for the entire data set. Processing the entire data set according to the training procedure is called a 'training epoch'. A neural network can be trained for any number of epochs, ranging from tens to thousands.

Therefore, deep neural networks can be used to model complex dependencies in a computationally optimal way, often requiring several orders of magnitude less computational time than exact methods. However, their application is not limited to regression problems. 

The neural networks have been implemented using the Flux.jl~\cite{Flux.jl-2018, innes:2018} library.

\section{Convolutional neural networks}

In practice, pure fully connected networks are rarely used on datasets where the existence of local features in the input data can be assumed (e.g. real-life photographs.) When dealing with a dataset where spatial coordinates can be assigned to each element of the input data, a fully connected layer cannot take advantage of the information about the relative position of pixels. Hence, such a layer assigns as much meaning to nonlocal features as to local ones -- but nonlocal features are more likely to be spurious, making the network prone to overfitting. The networks designed to remedy this problem are called convolutional neural networks (CNNs for short). The basic unit of a CNN is a convolutional layer constructed as described below.

Suppose the input image $\mathcal{I}$ is a 3-dimensional $W\times H\times C$ tensor, where $W$ is the image width, $H$ is the height, and $C$ is the number of channels (which can be, for example, the red, green, and blue channels of a photograph). A convolutional layer is defined by its convolutional filter, bias, and activation function. A convolutional filter $\mathcal{W}$ is then a 4-dimensional $d_1\times d_2\times C \times C_{out}$ tensor, where $d_1 \times d_2$ are the spatial dimensions of the receptive field of the filter (usually $d_1=d_2=d$), and $C_{out}$ is the number of channels in the output feature map, which can be arbitrarily high. This convolutional filter is applied in the following way:

\begin{equation}
    y_{ijk} = \sum\limits_{x=0}^{d_1-1}\sum\limits_{y=0}^{d_2-1}\sum\limits_{l=1}^{C} \mathcal{W}_{xykl}\cdot\mathbf{\mathcal{I}}_{i+x, j+y, l}
\end{equation}

After this, a $C$-dimensional bias $\mathbf{b}$ is added, and a nonlinear activation function $f$ is applied like it's done for fully connected neural networks. The output of a convolutional layer is then:

\begin{equation}
    a_{ijk} = f(y_{ijk} + b_k)
\end{equation}

Just like fully connected layers, the convolutional layers can be nested upon each other to extract increasingly complex and nonlocal features with each subsequent layer. 
%In most problems, the desired output can be represented as a vector rather than a 2D image -- for example, classification labels or reconstructed system parameters. To produce such an output, the feature map obtained after several convolutions is then input into a fully connected layer. In its simplest case, it is a simple linear transform from the feature map to the output. However, the fully connected part of a CNN can generally include several susequent fully connected layers.

In practice, convolutional layers are almost universally alternated with pooling layers which reduce the amount of computations required. A pooling layer with a receptive field $d\times d$ shrinks a $W\times H\times C$-sized image to a $W\div d \times H\div d \times C$-sized one by mapping each region of size $d\times d$ of the input image to a single pixel of the output image for each channel. 
%The most widespread variants are maximum pooling (each pixel of the output corresponds to the maximum value in its respective input region; used in this work), and mean pooling (same for the mean value).
Usually, the number of channels in convolutional filters is doubled after each maximum pooling operation to make the nonlocal features more expressive than local ones.

CNN's are also commonly regularized with batch normalization layers. For each channel $i$ of the input data, a batch normalization layer first normalizes the channel data to mean 0 and variance 1 across the entire data batch, then shifts it to have a new mean $\beta_i$ and variance $\gamma_i$, where $\boldsymbol{\beta}$ and $\boldsymbol{\gamma}$ are trainable parameters. Although the usefulness of batch normalization in stabilizing and accelerating deep neural network training is demonstrably significant~\cite{ioffe2015batch}, the reason for it is currently a disputed matter~\cite{santurkar2019does}.

\begin{comment}
In our work we have also used residual blocks. A residual block $\mathcal{R}$ processes its input by first applying a transform $\mathcal{F}$ to its input $\mathbf{x}$ (here, $\mathcal{F}$ must necessarily conserve the dimensionality of $\mathbf{x}$, i.e. map a $W\times H\times C$-sized image to a $W\times H\times C$-sized one), and then adding the result to the output: $\mathcal{R}(\mathbf{x}) = \mathcal{F}(\mathbf{x}) + \mathbf{x}$. The transform $\mathcal{F}$ can include any combination of the operations described above. 
In our work, a residual block consists of two subsequent convolutions, similar to the original paper~\cite{he2015deep}. Using residual blocks not only allows us to efficiently train deep neural networks by allowing the higher layers to access the output of the lower ones; it also makes our network design modular, allowing for more efficient transfer learning. 
By introducing untrained residual blocks to a network that previously had none, we expand its capacity, allowing it to be transferred to tasks more complex than the source problem.
\end{comment}

An architecture demonstrating the described principles can be seen in Fig.~3; see ~\cite{mehta2019high}
for detailed review.

In our work, we aimed to first train our neural network on a source problem (the Rice-Mele model dataset), and then retrain it for a target problem (the gapped graphene 
dataset). We found that a fully connected neural network 
was able to attain performance comparable to a CNN on the source problem. To do so, it required 
the input of the 2D frequency-phase spectra concatenated with their Fourier transforms along the phase axis. However, as FCNNs are prone to overfitting, they turned out to be ill-suited for retraining and performed poorly on the target problem. CNNs, on the other hand, can be retrained by only retraining the fully connected layers, without affecting the convolutions. Following this procedure, we attained good performance on both the source and the target problems (see figs.~4,~5).

\section{Input data}

The phase $\Delta\varphi$ was discretized into 32 steps. Per each laser cycle, 80 time points were recorded, allowing us to recover integer harmonics up to $N=40$, thus a single 2D plot has 41x32 points of the form $\log_{10}\left|j(N\omega, \Delta\varphi)\right|$. These were normalized to the mean value of 0 and standard deviation of 1. The normalized images were then presented as $41\times 32\times 1$ tensors, i.e. 2D images with a single channel.

We have generated data sets that consist of 65536 samples for pulses with no cubic phase, and 262144 samples for pulses with cubic phase. Next, $1\%$ of samples from each dataset have been set apart to form the test set. The required output were the band parameters, the CEP in the form of $(\cos \varphi, \sin \varphi)$, and the dimensionless chirp and cubic phase. 

During the training process, the data have undergone random circular shifts: before a training step, all 2D plots $\left|j(N\omega, \Delta\varphi)\right|$ in each batch 
were  transformed by a circular shift along the $\Delta\varphi$ axis. This has allowed us 
to train with smaller datasets and helped to combat overfitting.

\section{Source problem performance}

The average performance of the selected neural network on the test set of the Rice-Mele model dataset for different unknown parameters is given below. For parameters that do not change sign, the average relative error is also indicated.

\begin{figure}[H]
    \begin{tikzpicture}
        \node (t0) at (0, 0) {No phase noise, no multiplicative noise:};
        \node[anchor=north] (table0) at (t0.south) {\begin{tabular}{|c|c|c|c|c|c|}
        	\hline
        	&$\delta \epsilon_{min}$, eV&$\delta t_1$, eV&$\delta t_2$, eV&$\delta t_3$, eV & $\delta\varphi$, rad \\
        	\hline
        	abs&0.024& 0.030& 0.037& 0.025 &0.018
        	\\
        	\hline
        	rel&0.004& 0.012&&&\\
        	\hline
        \end{tabular}};
    
        \node[anchor=north] (t1) at (table0.south) {No phase noise, multiplicative noise 10\%:};
        \node[anchor=north] (table1) at (t1.south) {\begin{tabular}{|c|c|c|c|c|c|}
        	\hline
        	&$\delta \epsilon_{min}$, eV&$\delta t_1$, eV&$\delta t_2$, eV&$\delta t_3$, eV & $\delta\varphi$, rad \\
        	\hline
        	abs&0.026& 0.032& 0.040& 0.027 &0.018
        	\\
        	\hline
        	rel&0.005& 0.013&&&\\
        	\hline
        \end{tabular}};

        \node[anchor=north] (t50) at (table1.south) {Phase noise 50 mrad, multiplicative noise 10\%:};
        \node[anchor=north] (table50) at (t50.south) {\begin{tabular}{|c|c|c|c|c|c|}
        	\hline
        	&$\delta \epsilon_{min}$, eV&$\delta t_1$, eV&$\delta t_2$, eV&$\delta t_3$, eV & $\delta\varphi$, rad \\
        	\hline
        	abs&0.033& 0.073& 0.065& 0.035 &0.022
        	\\
        	\hline
        	rel&0.006& 0.021&&&\\
        	\hline
        \end{tabular}};
        
        \node[anchor=north] (t200) at (table50.south) {Phase noise 200 mrad, multiplicative noise 10\%:};
        \node[anchor=north] (table200) at (t200.south) {\begin{tabular}{|c|c|c|c|c|c|}
        	\hline
        	&$\delta \epsilon_{min}$, eV&$\delta t_1$, eV&$\delta t_2$, eV&$\delta t_3$, eV & $\delta\varphi$, rad \\
        	\hline
        	abs&0.046& 0.074& 0.089& 0.058 &0.047
        	\\
        	\hline
        	rel&0.008& 0.025&&&\\
        	\hline
        \end{tabular}};
        
        %\node[inner sep=0, anchor=north, yshift=-1cm] (pic1) at (table.south) {\includegraphics[width=\textwidth]{pictures/accuracy-0505_1536.png}};
        %\node[inner sep=0, anchor=north] (pic2) at (pic1.south)  {\includegraphics[width=\textwidth]{pictures/cep-0505_1536.png}};
    \end{tikzpicture}
    \caption{Average recovery performance for the CEP + 4 parameters Rice-Mele model dataset.}
\end{figure}
    
\begin{figure}[H]
    \begin{tikzpicture}
        \node[inner sep=0] (table) at (0, 0) {\begin{tabular}{|c|c|c|c|c|c|c|}
        	\hline
        	&$\delta \epsilon_{min}$, eV&$\delta t_1$, eV&$\delta t_2$, eV&$\delta t_3$, eV & $\delta\varphi$, rad & $\delta\beta$ \\
        	\hline
        	abs&0.043& 0.056& 0.062& 0.064&0.064&0.070
        	\\
        	\hline
        	rel&0.007& 0.026&&&&\\
        	\hline
        \end{tabular}};
        %\node[inner sep=0, anchor=north, yshift=-1cm] (pic1) at (table.south) {\includegraphics[width=\textwidth]{pictures/accuracy-0705_0839.png}};
        %\node[inner sep=0, anchor=north] (pic2) at (pic1.south)  {\includegraphics[width=\textwidth]{pictures/cep-0705_0839.png}};
    \end{tikzpicture}
    \caption{Recovery performance for the CEP + chirp + 4 parameters Rice-Mele model dataset.}
\end{figure}
    
\begin{figure}[H]
    \begin{tikzpicture}
        \node[inner sep=0] (table) at (0, 0) {\begin{tabular}{|c|c|c|c|c|c|c|c|}
        	\hline
        	&$\delta \epsilon_{min}$, eV&$\delta t_1$, eV&$\delta t_2$, eV&$\delta t_3$, eV& $\delta\varphi$, rad & $\delta\beta$ & $\delta\epsilon$ \\
        	\hline
        	abs&0.052& 0.069& 0.099& 0.066 &0.228&0.098&0.037
        	\\
        	\hline
        	rel&0.010& 0.026&&&&&\\
        	\hline
        \end{tabular}};
        %\node[inner sep=0, anchor=north, yshift=-1cm] (pic1) at (table.south) {\includegraphics[width=\textwidth]{pictures/accuracy-1005_1757.png}};
        %\node[inner sep=0, anchor=north] (pic2) at (pic1.south)  {\includegraphics[width=\textwidth]{pictures/cep-1005_1757.png}};
    \end{tikzpicture}
    \caption{Recovery performance for the CEP + chirp + cubic phase + 4 parameters Rice-Mele model dataset.}
\end{figure}

\section{Modeling high-harmonic response of gapped graphene}

To demonstrate that the concepts learned by the neural network on the original task can be applied to more difficult problems, we consider the case of hexagonal boron nitride (hBN), a two-dimensional material described by a Hamiltonian equivalent to gapped graphene:

\begin{equation}\label{eq:hamiltonian_hbn}
    \hat{H}_0(\mathbf{k}) = t_1 \sum\limits_i \left[\sigma_x \cos(\mathbf{k}\cdot\mathbf{a}_i) - \sigma_y \sin(\mathbf{k}\cdot\mathbf{a}_i)\right] + m\sigma_z
\end{equation}

The parameters of this material are taken within the following ranges: $m\in [4.0, 7.27]$ eV, $t_1\in [0.08, 0.16]$ a.u. (atomic units; $1\text{ a.u.} = 27.2\text{ eV}$). The incident laser field is a single-cycle pulse with wavelength $\lambda = 3$ $\mu m$, peak power $P = 8.5\cdot 10^{12}$ $W/cm^2$. The pulse width, carrier frequency, and peak power are kept constant for all samples. The decoherence time $T_2$ set to 2 fs.

We model the high-harmonic response by solving the density matrix master equation (atomic units are used),
\begin{equation}
\partial_t \rho_{nm} (\mathbf{k},t) = -i [H(\mathbf{k},t),\rho(\mathbf{k},t)]_{nm} - \frac{(1-\delta_{nm}\rho_{nm}(\mathbf{k},t))}{T_2},
\end{equation}
using the code introduced in~\cite{silva2019high, silva2019topological}. The initial state is a mixed state with no coherence between the eigenstates, and given by the fully-filled valence band. The time-dependent interaction is performed in the length gauge and in the dipole approximation,
\begin{equation}
H (\mathbf{k},t) = H_0(\mathbf{k}) + E(t)\cdot \mathbf{r},
\end{equation}
where $H_0(\mathbf{k})$ is the field-free Hamiltonian in Eq.~\ref{eq:hamiltonian_hbn} and $E(t)$ is the time-dependent electric field. The position operator is given by $\mathbf{r} = i \partial_{\mathbf{k}} + A(\mathbf{k})$, where $A(\mathbf{k})$ is the Berry connection~\cite{blount1962formalisms}. The current is then obtained as
\begin{equation}
\mathbf{J}(t) = - \sum_{\mathbf{k}} \text{Tr}(\mathbf{v}_{\mathbf{k}} \rho(\mathbf{k},t))/N_{\mathbf{k}},
\end{equation}
where $N_{\mathbf{k}}$ are the number of $k$-points, and $\mathbf{v}_{\mathbf{k}} = -i [\mathbf{r},H_0(\mathbf{k}) + E(t)\cdot \mathbf{r}] = -i[\mathbf{r},H_0(\mathbf{k})]$ is the velocity operator.

%We model the high-harmonic response by integrating the semiconductor Bloch equations:
%\begin{equation}
%\begin{split}
%\dot{\pi}(\mathbf{k}, t) & = -\frac{\pi}{T_2} - i\Omega(\mathbf{k}, t)w(\mathbf{k}, t)\exp(-iS(\mathbf{k}, t))\\
%\dot{n}_m(\mathbf{k}, t) &= is_m\Omega^*(\mathbf{k}, t)\pi(\mathbf{k}, t)\exp(iS(\mathbf{k}, t)) + \text{c.c.}
%\end{split}
%\end{equation}
%where $\pi(\mathbf{k}, t)$ is the interband coherence, $n_m(\mathbf{k}, t)$ is the population of band $m$ for $m=\{c, v\}$, $w(\mathbf{k}, t) = n_c-n_v$, $\Omega(\mathbf{k}, t) = \mathbf{F}(t)\cdot \mathbf{d}(\mathbf{k}+\mathbf{A}(t))$, $S(\mathbf{k}, t) = \int_{-\infty}^t dt' \epsilon_g(\mathbf{k}+\mathbf{A}(t'))$, $s_m = \{-1,1\}$. The inter- and intraband current are then computed as:
%
%\begin{equation}
%    \mathbf{p}(\mathbf{k}, t) = \pi(\mathbf{k}, t)\exp(iS(\mathbf{k}, t))\mathbf{d}(\mathbf{k}+\mathbf{A}(t))
%\end{equation}
%
%\begin{equation}
%    \mathbf{j}_{er}(t) = \frac{d}{dt}\int  \mathbf{p}(\mathbf{k}, t)d\mathbf{k}
%\end{equation}
%
%\begin{equation}
%    v_m(\mathbf{k}) = \nabla_\mathbf{k}\epsilon_m(\mathbf{k})
%\end{equation}
%
%\begin{equation}
%    \mathbf{j}_{er}(t) = \frac{d}{dt}\sum\limits_m\int  \mathbf{v}_m(\mathbf{k}+\mathbf{A}(t))n_m(\mathbf{k}, t)d\mathbf{k}
%\end{equation}
%
%\begin{equation}
%    \mathbf{j}(t) = \mathbf{j}_{er}(t) + \mathbf{j}_{ra}(t)
%\end{equation}

We checked the high harmonic spectrum was converged for a $k$-grid of $n_{k_x} = n_{k_y} = 200$ points and a time step of $\Delta t = 0.8$~a.u. Same as before, for each crystal sample, consisting of a mass, first neighbor hopping, and initial CEP, we measure the current between H0 and H40. 

\section{Transfer learning}

The method of transfer learning is applied to problems for which there isn't enough data to train a neural network (commonly named "target" problems), but which have a similar problem ("source" problem) with sufficient training data available. The first stage consists of training the model (in our case, a neural network) to solve the source problem. The deeper layers of our network learn more abstract representations of the input data. However, as the information flow approaches the output, the concepts learned by the layers become less abstract and more specialized on the exact problem. Thus, in the second stage, the neural network is partially retrained for the target problem, with only the last layers being retrained. The retrained network may or may not be used as part of a larger neural network which modifies its inputs or uses the outputs of its hidden layers.

Alternatively, transfer learning may be used for two similar problems even when there is sufficient data for the target probkem. A network pre-trained on a source problem may be retrained completely for the target, but its pre-trained parameters still prove to be a better initial guess than random initialization.

We thus use the idea of transfer learning in two senses. First, for the source (TDSE) problem, networks trained on datasets with less unknown nonlinear phase parameters are used as initial guesses to train networks to recover more complex spectral phases -- i.e., the networks designed to recover the CEP and band parameters for unchirped pulses are retrained to process chirped pulses, after which they are retrained on the dataset with a cubic phase.  

We also use transfer learning to train our neural network to recover pulses and band parameters from gapped graphene spectra. After pretraining it on the source dataset we use 1280 gapped graphene responses as target dataset. Due to its small size, we use $20\%$ of the dataset for validation. We augment the target data in the same way as we do with the source data. 

For both networks (band parameter and CEP), we retrain the entire network.

\section{Retrained network performance}

The average performance of our neural networks on the noisy gapped graphene dataset after retraining is given below.

\begin{figure}[H]
    \begin{tikzpicture}
        \node[inner sep=0] (table) at (0, 0) {\begin{tabular}{|c|c|c|c|}
        	\hline
        	&$\delta m$, eV&$\delta t_1$, a.u. & $\delta\varphi$, rad \\
        	\hline
        	abs&0.021& $5.3\cdot 10^{-4}$& 0.0184
        	\\
        	\hline
        	rel&$3.9\cdot 10^{-3}$& $4.5\cdot 10^{-3}$&\\
        	\hline
        \end{tabular}};
        %\node[inner sep=0, anchor=north, yshift=-1cm] (pic1) at (table.south) {\includegraphics[width=\textwidth]{pictures/accuracy-0605_1337.png}};
        %\node[inner sep=0, anchor=north, yshift=-0.5cm] (pic2) at (pic1.south)  {\includegraphics[width=\textwidth]{pictures/cep-0605_1337.png}};
    \end{tikzpicture}

    \caption{Average error of the neural network recovery for the gapped graphene dataset. a.u. designates 1 atomic unit.}
\end{figure}

% Bibliography
\bibliography{refs}

%Manual citation list
%\begin{thebibliography}{1}
%\bibitem{Zhang:14}
%Y.~Zhang, S.~Qiao, L.~Sun, Q.~W. Shi, W.~Huang, %L.~Li, and Z.~Yang,
 % \enquote{Photoinduced active terahertz metamaterials with nanostructured
  %vanadium dioxide film deposited by sol-gel method,} Opt. Express \textbf{22},
  %11070--11078 (2014).
%\end{thebibliography}